\documentclass[preprint,aps,twocolumn]{revtex4}%
\usepackage{amsfonts}
\usepackage{amsmath}
\usepackage{amssymb}
\usepackage{graphicx}%
\begin{document}
\title{Exact calculation of the number of degrees of freedom of a rigid body
constituted by \textit{n }particles}
\author{Jorge Bernal}
\email{jorge.bernal@dacb.ujat.mx}
\affiliation{Universidad Ju\'{a}rez Aut\'{o}noma de Tabasco, km. 1 Carretera
Cunduac\'{a}n-Jalpa, Cunduac\'{a}n, Tabasco, M\'{e}xico, C.P. 86690}
\author{Roberto Flowers-Cano}
\email{flowerscano@hotmail.com}
\affiliation{Universidad Ju\'{a}rez Aut\'{o}noma de Tabasco, km. 1 Carretera
Cunduac\'{a}n-Jalpa, Cunduac\'{a}n, Tabasco, M\'{e}xico, C.P. 86690}
\author{Adrian Carbajal-Dominguez}
\email{adrian.carbajal@dacb.ujat.mx}
\affiliation{Universidad Ju\'{a}rez Aut\'{o}noma de Tabasco, km. 1 Carretera
Cunduac\'{a}n-Jalpa, Cunduac\'{a}n, Tabasco, M\'{e}xico, C.P. 86690}

\begin{abstract}
In this work we correct a calculation made by Albert Einstein that appears in
his book titled "The Meaning of Relativity" (Princeton, 1953), and by means of
which he tries to obtain the number of degrees of freedom of a system
constituted by \textit{n} particles with fixed relative distances and which
are immerse in a three-dimensional space. As a result of our analysis, we
develop expressions which yield the number of degrees of freedom of an
analogous system, not only in three, but in any arbitrary number \textit{D} of dimensions.

\end{abstract}

\maketitle

\bigskip

The number of independent coordinate variables needed to simultaneously
determine the position of every particle in a dynamical system is called the
number of degrees of freedom of that system. So a system of $n$ free particles
in a three-dimensional space has $3n$\textit{ }degrees of freedom, because
three coordinates are needed to specify the location of the center of mass of
each particle. However, if the particles are no longer all free, but there are
restrictions imposed on the system, the number of degrees of freedom will be
less than $3n$; $3n$ coordinates are still needed to locate the centers of
mass, but less than $3n$ values are assignable at will to the coordinate
variables \cite{Fowles}. Specifically, we are interested in the system made up
of $n$\textit{ }particles in three-dimensional space, which hold fixed
distances between them. In the sake of clarity, this system will be referred
to from now on as $S_{3}$, and the number of its degrees of freedom will be
referred to as $N_{3}$.

Usually, $N_{3}$ is calculated by giving $S_{3}$ the treatment of a rigid
body. Mechanics recognizes two types of rigid bodies: the ones made up by a
continuous distribution of mass; and those formed by $n$ mass points joined by
rigid links \cite{Scheck}. Thus, $S_{3}$ is equivalent to a rigid body of the
second type.

It is not difficult to calculate the number of degrees of freedom of a rigid
body of continuous mass. For most cases, the number of degrees of freedom is
six, as three coordinates are needed to locate the body%
\'{}%
s center of mass and three more to describe its orientation \cite{Fowles}%
,\cite{Scheck}. But if the mass is all distributed along a single line, then
it will be impossible for the body to rotate about that line, and therefore,
such a body has only five degrees of freedom \cite{Scheck},\cite{Goldstein}. A
similar reasoning is used to calculate $N_{3}$, after assuming that $S_{3}$
may be viewed as a sole body instead of a collection of particles. Hence,
$N_{3}$ is five when $n=2$, since the mass points lie all along the same line,
and is six when $n>2$ \cite{Barford}. The case in which $n>2$ particles lie
all on the same line will not be considered in this work.

This same results should be attainable through individual consideration of the
particle which make up $S_{3}$. Counting the number of degrees of freedom of
$S_{3}$ is fairly easy when $n$ is equal to two: six are the coordinates
needed to locate the centers of mass of the particles, but there is one
restriction (one rigid link), so the number of degrees of freedom of $S_{3}$
is five. It is not hard either to calculate the number of degrees of freedom
of $S_{3}$ when $n=3$. Then, nine coordinates are needed to specify the
positions of the particles%
\'{}%
centers of mass, but since there are three restrictions, the number of degrees
of freedom is six. That is, if the triad does not lie all along the same line;
if that is so, there are four restricitions and the number of degrees of
freedom of the system is again five.

The operation of calculating $N_{3}$ by consideration of the individual
particles would be much easier if an expression which would yield the number
of degrees of freedom of $S_{3}$ for any given value of $n$ was developed.
Albert Einstein figures among those who tried to develop an expression such.
Einstein dealt with this problem in one of his books \cite{Einstein}, using it
as an example of the importance that geometrical concepts have a
correspondence with real objects. He reasoned more or less along the following lines:

If \ one particle (let this particle be called particle 1), is arbitrarily
chosen from among the $n$ that compose $S_{3}$, $n-1$ equations are needed to
express the fact that this particle holds fixed distances with the rest%
\begin{equation}
\left(  x_{j}-x_{1}\right)  ^{2}+\left(  y_{j}-y_{1}\right)  ^{2}+\left(
z_{j}-z_{1}\right)  ^{2}=d \label{ceq}%
\end{equation}%
\[
\text{where }d\text{ is a constant and }j=1,2,3,...,n
\]

But when a second particle is taken into consideration, to express that the
distances between this and the other particles remain constant, only $n-2$
equations are needed, because the equation that shows that the distance
between particles 1 and 2 is constant is already included in (\ref{ceq}). If a
third particle is considered, there would be $n-3$ equations more; for a
fourth particle, there would be $n-4$ equations more, and so on. In total,
there are $\frac{n\left(  n-1\right)  }{2}\ $different equations. These
equations represent the system%
\'{}%
s restrictions; they are the constraint equations of the system.

Einstein must have thought that he would obtain the number of degrees of
freedom of $S_{3}$\ merely by substracting the number of constraint equations
from $3n:$%
\begin{equation}
N_{3}=3n-\frac{n\left(  n-1\right)  }{2} \label{N3eqn1}%
\end{equation}

If (\ref{N3eqn1}) is solved for $n>4$, it will be seen that the values of
$N_{3}$ differ from those obtained when $S_{3}$ was viewed as a single body.
Why does this happen? Maybe because it is not all appropiate to consider the
collection of particles with rigid links as one body. Or more likely, because
the count of the degrees of freedom of $S_{3}$ by consideration of the
individual particles was not done correctly. Which ever the reason may be, we
will soon find out.

As it turns out, there is something definitely wrong with (\ref{N3eqn1}), and
it is that%
\begin{equation}
3n-\frac{n\left(  n-1\right)  }{2}\approx-\frac{n^{2}}{2}<0, \label{aprox}%
\end{equation}%
\[
\text{for}\ n>>1,
\]

which is absurd.

Einstein did notice this flaw, because in his book, instead of (\ref{N3eqn1})
he has:%
\begin{equation}
N_{3}=\frac{n\left(  n-1\right)  }{2}-3n \label{N3eqn2}%
\end{equation}

We cannot think of any physical or mathematical justification for this change
of signs, and although it removes the problem of getting a negative value of
$N_{3}$ when $n>>1$, it brings up a new problem.

In the limit when $n$ tends to infinity, the system $S_{3}$ is equivalent to a
rigid body of continuous mass. So it would be expected that if the limit of
$N_{3}$ is taken when $n$ tends to infinity, this limit should be equal to
six. But this does not hold true for $N_{3}$ as defined in (\ref{N3eqn2}); the
limit when $n$ tends to infinity diverges.

Einstein introduced, as a footnote, the following correction:%
\begin{equation}
N_{3}=\frac{n(n-1)}{2}-3n+6 \label{N3eqn3}%
\end{equation}

Nonetheless, the limit when $n$ tends to infinity of the modified $N_{3}$ is
still undefined, so (\ref{N3eqn3}) cannot be the correct expression for
$N_{3}$ either.

When we took up the task of developing an accurate expression for $N_{3}$, we
did not take off from where Einstein left the problem, but instead, we
directed our attentions back to (\ref{N3eqn1}), which is the expression that
Einstein must have come up with originally, in spite of the fact that it doesn%
\'{}%
t appear in his book. We did so because, as incorrect as it may be, there is a
consistent line of thinking behind expression (\ref{N3eqn1}), which there is
not behind expressions (\ref{N3eqn2}) or (\ref{N3eqn3}).

Expression (\ref{aprox}) gave us a hint of where the flaw in (\ref{N3eqn1})
may be. Not in the signs, but rather, in the lack of a term. A term that
shouldn%
\'{}%
t be a constant, but dependent of $n$. A term that added up to the other two
would not only make $N_{3}$ possitive for $n>>1$, but actually equal to six.
So there must be an aditional source of degrees of freedom which Einstein
missed to consider. If we could identify where this source of degrees of
freedom was, we would have our problem solved.

A group of $n$ particles may rotate in space without dissatisfying the
condition that the distances between the particles remain constant. However,
it is meanigless to talk about rotations without first establishing an
adequate reference frame. To do so we arbitrarily selected three particles
from $S_{3}$; the points were the centers of mass of these particles are
located generate a plane $P$ in three-dimensional space. And the vector
$\mathbf{v}$, which is orthogonal to $P$, designates an arbitrary direction in
space. We must point out that we are defining $\mathbf{v}$ as a fixed vector,
and that it is perpendicular to $P$ in its original position, but as
$S_{3\text{ }}$rotates, this perpendicularity relation will be lost.
Therefore, it is convenient to make a copy of $P$, which we will call $P~%
\acute{}%
$, and hold this copy fixed in the original position of $P$. Thus $\mathbf{v}$
will allways be orthogonal to $P~%
\acute{}%
$. By considering the plane $P~%
\acute{}%
$ and its normal vector, we are defining a three-dimensional coordinate system.

Now, if we choose two particles, different from the ones used to generate the
plane, the line that joins their centers of mass is a possible rotation axis
for $S_{3}$. And since the number of ways in which pairs may be chosen from a
set of $n-3$ particles is%
\begin{equation}
{\small C}_{n-3}^{2}{\small =}\frac{(n-3)!}{2!(n-5)!}{\small =}\frac
{(n-3)(n-4)}{2}{\small ,} \label{comb1}%
\end{equation}%
\[
\text{for }n\geq3.
\]

There will be an equal number of such axes. Each of this axes forms with the
direction of the vector $\mathbf{v}$ an angle $\varphi_{i}$ which is a
function of time and determines a possible rotation of the system. In general,
the different $\varphi_{i}$ will not hold relations of linear independence.

We believe that the number of $\varphi_{i}$ allowed to $S_{3}$ for a given
value of $n$ is the term missing in Einstein's calculation, and we propose
that the number of degrees of freedom for the system $S_{3}$ is given by:%
\begin{equation}
{\small N}_{3}{\small =3n-}\frac{n(n-1)}{2}{\small +}\frac{(n-3)(n-4)}%
{2}{\small =6,} \label{N3eqn4}%
\end{equation}%
\[
\text{when }n\geq3.
\]

However, (\ref{N3eqn1}) seems to be the correct expression for $n=2$. It also
works for $n=3$ and $n=4$, which is not surprising, since for this value of
$n$ the last term in expression (\ref{N3eqn4}) is equal to zero, so
(\ref{N3eqn4}) and (\ref{N3eqn1}) are equivalent.

Once we had developed this expressions, we were curious on wether, by
following the same line of reasoning, we could calculate the number of degrees
of freedom of $S_{4}$, that is, of the system made up by $n$ particles with
fixed relative distances, but which is, unlike $S_{3}$, immerse in a four
dimentional space.

In this four-dimentional case, four coordinates are needed to locate the
center of mass of each particle, which makes $4n$ coordinates for the set of
$n$ particles. And the number of constraint equations is the same as for
$S_{3}$

In principle, the number of degrees of freedom should be the same as for a
tetra-dimensional rigid body. And in four dimensions there are ten degrees of
freedom for the rigid body: four coordinates are needed to locate its center
of mass and there are six possible rotation angles. Now, in the case of the
$n$ particles with fixed distances, we need $4n$ coordinates to locate the
particles%
\'{}
centers of mass, while the number of distances is still $\frac{n(n-1)}{2}.$
And the number of possible rotation angles is obtain observing that a
"hiperplane" can be defined with four points and that the number of diferent
ways in which pairs may be chosen from a group of $n-4$ particles is given by:%
\begin{equation}
{\small C}_{n-4}^{2}{\small =}\frac{(n-4)!}{2!(n-6)!}{\small =}\frac
{(n-4)(n-5)}{2}{\small ,} \label{comb2}%
\end{equation}%
\[
\text{for }n\geq4.
\]

Then, the number of degrees of freedom of $S_{4}$ is%
\begin{equation}
{\small N}_{4}{\small =4n-}\frac{n(n-1)}{2}{\small +}\frac{(n-4)(n-5)}%
{2}{\small =10,} \label{N4eqn1}%
\end{equation}%
\[
\text{when }n\geq4,
\]

and%
\begin{equation}
N_{4}=4n-\frac{n(n-1)}{2}, \label{N4eqn2}%
\end{equation}%
\[
\text{when }2\leq n\leq5,
\]

since the number of possible $\varphi_{i}$ is equal to zero for these values
of $n.$

That $N_{4}$ is equal to ten for any value of $n$ less than or equal to four
is consistent with the fact that ten is also the number of degrees of freedom
of a rigid body in four-dimentional space (four coordinates are needed to
locate the center of mass, and six more to describe the orientation of the
body. Indeed, our procedure works for the four-dimentional as it does for the
three-dimensional case. Moreover, we believe that it works for the general
case. We propose that for a system of $n$ particles with fixed relative
distances, immerse in a space of $D$ dimensions, the number of degrees of
freedom is given by:%
\begin{equation}
{\small N}_{D}{\small =Dn-}\frac{n(n-1)}{2}{\small +}\frac{(n-D)(n-D-1)}{2}
\label{NDeqn1}%
\end{equation}%
\[
{\small =}\frac{D(D+1)}{2}{\small ,}\text{when }n\geq D,
\]

and by:%
\begin{equation}
N_{D}=Dn-\frac{n(n-1)}{2}, \label{NDeqn2}%
\end{equation}%
\[
\text{when }2\leq n\leq D+1.
\]

These results coincide entirely with those which would have been obtained by
viewing $S_{D}$ as a single body.

Counting the number of degrees of freedom of $S_{D}$ by consideration of the
individual particles is something which had never been done before. Just the
three-dimensional case proved to be complicated enough. Even for Albert
Einstein, who was never able to write the correct expressions for the number
of degrees of freedom of $S_{3}$ in \cite{Einstein}, in spite of several
revisions he made of this book.

There seemed to be contradictions between the values of $N_{3}$ obtained
viewing $S_{3}$ as a sole body and those reached by considering the individual
particles. This was only because the count of the degrees of freedom of
$S_{3}$ from the latter standpoint was never done properly. In this paper, we
prove that both methods are equivalent, not only in three, but in any number
$D$ of dimensions.

This may be of interest for those who study the Kinetic Theory of Gases. In
the Kinetic Theory of Gases and more specifically, in the Ideal Gas Model, the
internal energy and the heat capacities at constant volume and constant
pressure of an ideal gas are calculated as functions of the degrees of freedom
of the gas, which are counted per molecule. And for molecules consisting of
more than one atom, the number of degrees of freedom is calculated treating
the molecules as rigid bodies. Thus, a diatomic molecule has five degrees of
freedom and a polyatomic molecule has six. According to the Equipartition of
Energy Theorem, each of these degrees of freedom is associated to an energy of
quantity $\frac{1}{2}kT$. Hence, the internal energy $U$ of a diatomic
molecule is $U=\frac{5}{2}kT$ and that of a polyatomic molecule is $U=3kT$.
Multiplying these results by Avogadro%
\'{}%
s number, $N_{A}=6\times10^{23}$, gives the internal energy of an ideal gas,
which is $U=\frac{5}{2}N_{A}kT=\frac{5}{2}RT$ and $U=3N_{A}kT=3RT$ for
diatomic and polyatomic gases, respectively \cite{Resnick},\cite{Castellan}.

The heat capacity at constant volume $C_{v}$ is related to the internal energy
by the expression $C_{v}=\left(  \frac{\partial U}{\partial T}\right)  $, thus
$C_{v}=\frac{5}{2}R$ for diatomic gases and $C_{v}=3R$ for the polyatomic
ones. The heat capacity at constant pressure $C_{p}$ is given by $C_{p}%
=C_{v}+R$.

The values of the heat capacities predicted using the Ideal Gas Model agree
very well with the values obtained experimentally in the case of diatomic
gases, but fall rather short for polyatomic gases \cite{Resnick},
\cite{Castellan}. This is due to the fact that besides the energies associated
with the traslational and rotational degrees of freedom, there is also
vibrational energy. This vibrational energy is quanticized, which means that
it does not spread over a continuous spectrum of values, but is distributed in
discrete states \cite{Castellan}, \cite{Mandl}.

In the case of most diatomic molecules, the difference between the state of
lowest energy (the ground state) and the state that follows is such, that the
leap from the ground state to the next may only be achieved at temperatures of
approximately 3500 K. Thus, at room temperatures, the vibrational energy will
remain in the ground state and its contributions to the total internal energy
of the molecule is negligible. Something very different occurs with polyatomic
gases, where the molecules have several independent vibration modes. For some
of this modes, the spacing between energy states is considerably smaller than
for diatomic molecules. Hence, the vibrational energy will make an important
contribution to the total internal energy of a polyatomic molecule at room
temperature, or even less. Once the vibrational energy is considered, the
predicted heat capacities have a very good correspondence with experimental
values \cite{Mandl},\cite{Hill}.

Anyhow, the aditional consideration of this quanticized vibrational energy
does not modify the fact that the rotational and traslational energies of a
gas molecule are calculated by treating this molecule as a rigid body.
Treating molecules as rigid bodies is correct, but it had never been formally
justified. This work gives a formal justification to this procedure.

Furthermore, we believe that this paper clarifies the so-called "degree of
freedom paradox". This paradox consists in that, if we make a microscopical
analysis of a system which treated as a rigid body has a finite number of
degree of freedom, it turns out that it has an infinite number of degrees of
freedom and therefore, infinite heat capacities, which is absurd \cite{Huang}.
This contradiction was attributed to a flaw in classical mechanics. Our work
suggests that rather, it is a result of not knowing how to count the number of
degrees of freedom particle by particle.

This work may also imply that statements like the following are not correct.
According to Herbert Goldstein, "a rigid body with $N$ particles can at most
have $3N$ degrees of freedom", as can be read in his Classical Mechanics
textbook \cite{Goldstein}, in the chapter dealing with the kinematics of rigid
body motion. However, our analysis shows that the maximum number of degrees of
freedom for any rigid body in three dimensional space is six.

In conclusion, we obtained expression which yield the number of degrees of
freedom of a rigid body constituted by \textit{n }particles in a
three-dimensional space and we extended our results to an arbitrary number
\textit{D }of spatial dimentions. The results for the three-dimensional case
disagree with those obtained by Albert Einstein and which appear in
\cite{Einstein}. We believe that with our analysis of the three-dimentional
case we can justify, formally, that a rigid non-linear polyatomic molecule
allways has six degrees of freedom, situation which has not been sufficiently
explained in literature, in spite of its widespread use in the calculation of
the internal energies and heat capacities of ideal polyatomic gases.

\textit{We thank Trinidad Cruz-S\'{a}nchez for his valuable contribution to
the fulfillment of this work.}

\bigskip

\newpage

Figure Captions

\bigskip

Figure 1. At instant $t=0$ (\textbf{a}) the system is in its initial
posistion. The line that conects the centers of mass of two arbitrary
particles forms an angle $\varphi(t=0)$ with the direction of the vector
$\mathbf{v}$ orthogonal to the reference plane $P$. At a future instant $t=t~%
\acute{}%
$ (\textbf{b}) the system has rotated respect to its original position. The
plane $P$ has moved, but a copy $P~%
\acute{}%
$ remains in the original position of $P$, so now $\mathbf{v}$ is
perpendicular to $P~%
\acute{}%
$. And the line that joins the centers of mass of the particles we had
considered forms an angle $\varphi(t~%
\acute{}%
)$ with the direction of $\mathbf{v}$.

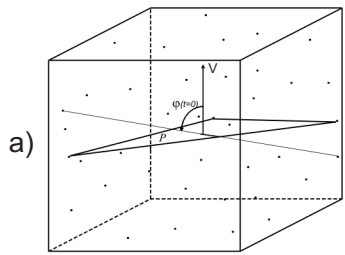
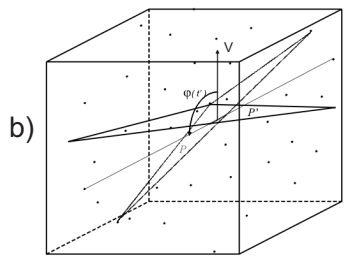
\end{document}